\documentclass[12pt]{article}
\usepackage[pdftex]{graphicx}
\usepackage{caption}
\usepackage{subcaption}
\usepackage{dcolumn}% Align table columns on decimal point
\usepackage{bm}% bold math
\usepackage{color}
\usepackage{amsfonts}
\usepackage{amsmath}
\usepackage{stmaryrd}
\usepackage{epsfig}
\usepackage[english]{babel}
\usepackage[all]{xy}

 \newtheorem{prop}{Proposition}

\usepackage[top=25.4mm, bottom=25.4mm, left=31.7mm, right=32.2mm]{geometry}
\begin{document}

\renewcommand{\baselinestretch}{1.5}

\title{Supersymmetric Sawada-Kotera Equation: B\"{a}cklund-Darboux Transformations and Applications}

\author{Hui Mao$^\dagger$, Q. P. Liu$^\dagger$\footnote{Correspondence author.Email: qpl@cumtb.edu.cn} and Lingling Xue$^\ddagger$
\\
$^\dagger$Department of Mathematics,\\
China University of Mining and Technology,\\
Beijing 100083, P. R. China\\
$^\ddagger$Department of Mathematics,\\
Ningbo University, Ningbo 315211, P. R. China}

\date{}
\maketitle
%\date{\today}

\begin{abstract}
In this paper, we construct a Darboux transformation and the related B\"acklund transformation for the supersymmetric Sawada-Kotera (SSK) equation. The associated nonlinear superposition formula is also worked out. We demonstrate that these are natural extensions of the  similar results of the Sawada-Kotera equation and may be applied to produce the solutions of the SSK equation. Also, we  present
two semi-discrete systems and show that the continuum limit of one of them  goes to the SKK equation.
\end{abstract}

{\bf Key words:} {solitons, integrable system, supersymmetry, B\"acklund transformation, Darboux transformation, nonlinear superposition formula, discrete integrable system}
\maketitle %\maketitle must follow title, authors, abstract and

\newpage

\section{Introduction}
In searching for the Korteweg-de Vries type equations with $N-$soliton solutions, Sawada and Kotera \cite{sk}, also Caudrey, Dodd and Gibbon \cite{caudrey-dg} independently, found the following fifth order evolution equation
\begin{equation}
u_t+u_{xxxxx}+5uu_{xxx}+5u_{x}u_{xx}+5u^2u_{x}=0.
\label{SK}
\end{equation}
This equation, known as the SK equation or CDGSK equation, has been one of the most important equations in the soliton theory and a large amount of results have been accumulated for it.
Satsuma and Kaup \cite{satsuma-k}, within the framework of Hirota bilinear method, obtained its B\"acklund transformations, Lax pair and infinitely many conserved quantities.   By means of the prolongation theory, Dodd and Gibbon worked out  the similar results \cite{dodd-g}. Fordy and Gibbons \cite{fordy-g}, independently Hirota and Ramani \cite{hirota-r}, shown that the SK equation is associated with another fifth order evolution equation, namely Kaup-Kupershmidt equation, and in particular these two systems  share a common modification \cite{fordy-g}. Kaup developed the inverse scattering method to the SK equation \cite{kaup}. Fuchssteiner and Oevel brought the SK equation into the bi-Hamiltonian formulation \cite{fuchssteiner-o}. According to Date {\it et al}, SK equation is a particular flow of the BKP hierarchy \cite{date-jkm1,date-jkm2}.  Levi and Ragnisco constructed the Darboux transformation for SK equation \cite{levi-r} (see also \cite{nimmo, athorne-n}) and  a nonlinear superposition formula was found by Hu and Li \cite{hu-l}.
Most recently, Geng, He and Wu constructed the algebro-geometric solutions for the SK hierarchy \cite{geng-hw}. For more  results and properties of the SK equation, one is referred to    \cite{aiyer-fo,euler,fordy,hirota1, hirota2,parker,lou, rogers-c,weiss, musette} and the references there.

With Tian, one of the authors proposed a  supersymmetric SK equation  \cite{TL}, which reads as
\begin{equation}
\phi_t+\phi_{xxxxx}+5\phi_{xxx}\phi'+5\phi_{xx}\phi'_x+5\phi_{x}\phi'^2=0,
\label{SSK}
\end{equation}
where $\phi=\phi(x,t,\theta)$ is a super fermionic function depending on temporal variable $t$, spatial variable $x$ and its fermionic counterpart $\theta$. ${\cal D}$ denotes the super derivative defined by ${\cal D}=\partial_{\theta}+\theta\partial_x$. For simplicity, here and in the sequel, we denote super derivative by prime and usual derivative with respect to $x$ by subscript $x$. To see the connection with the SK equation \eqref{SK}, we assume $\phi=\theta u(x,t)+\xi(x,t),$
where $u=u(x,t)$ is a bosonic (even) function while
$\xi=\xi(x,t)$ is a fermionic (odd) one, then
the SSK equation (\ref{SSK}) in components takes the following form
\begin{subequations}\label{cSSK}
 \begin{align}
  u_t+u_{xxxxx}+5uu_{xxx}+5u_{x}u_{xx}+5u^2u_{x}-5\xi_{xxx}\xi_{x}&=0,\\
  \xi_t+\xi_{xxxxx}+5u\xi_{xxx}+5u_{x}\xi_{xx}+5u\xi_{x}&=0,
 \end{align}
\end{subequations}
which reduces  to \eqref{SK} when the fermionic variable $\xi$ is set to zero. It is mentioned that the SSK equation also appears in the symmetry classification of supersymmetric integrable systems \cite{Tian-Wang}.
%the follow SK equation when $\xi=0$,
%\begin{equation}
%u_t+u_{xxxxx}+5uu_{xxx}+5u_{x}u_{xx}+5u^2u_{x}%=0.
%\label{SK}
%\end{equation}

As its classical counterpart, the SSK equation is also integrable and its integrability is ensured by presenting a Lax representation, the  existence of infinitely many conserved quantities and a recursion operator \cite{TL}. It is interesting to note the SSK equation possesses odd Hamiltonian structures and is a bi-Hamiltonian system \cite{popowicz}. Subsequent works show that the SSK equation is  associated with supersymmetric Kawamoto equation \cite{liu-pt} and passes the Painlev\'{e} test \cite{li-zl}.

%To the best of our knowledge, Hirota's bilinear %representation and Darboux transformations and %soliton solution of the SSK equation still %haven't been found. Also, any form of B\"acklund %transformation is not known to this equation.

The purpose of this paper is to construct a Darboux transformation and the related B\"{a}cklund  transformation for the SSK equation and study their applications. While B\"{a}cklund transformations have their origins from differential geometry (see \cite{Gu, rogers} and the references there), it is well known that
Darboux and  B\"{a}cklund transformations play a vital role in the study of nonlinear systems and the related theory constitutes an integrated part of the soliton theory (see\cite{Gu,rogers,matveev,cies,dl} for example).
B\"{a}cklund transformations have been known to be an effective approach to construction of solutions for nonlinear systems, furthermore they may be applied to generate new integrable systems, both continuous and discrete \cite{levi1980,levi1981,frank}. It is remarked that the applications of B\"acklund transformations to integrable discretization of super or supersymmetric integrable systems were developed only recently\cite{sasha,zhou, xll,xl,xue,carstea2015, mao-l}.
%Very recently, the relationship between these %results and extensions of Yang-Baxter map was %explored \cite{sasha1, sasha2}.

The  paper is arranged  as follows. In  next section,  we recall the Lax pair for the SSK equation and construct its  Darboux and B\"{a}cklund transformations. As  a simple application, 1-soliton solution to the equation is obtained  from the associated  Darboux transformation. In section 3,  we present a nonlinear superposition formula for the SSK equation and a 2-soliton solution is worked out. Then in the last section, we relate  the obtained B\"{a}cklund  transformation and nonlinear superposition formula to super differential-difference integrable systems. In particular, by taking continuum limit we show that one of the systems gives the SSK equation.

%Finally, we point out that all the results in %this article can be used in SK %equation(\ref{SK}), If let the  fermionic part %to be zero.

\section{B\"{a}cklund-Darboux Transformations}
To construct a Darboux transformation for the SKK equation \eqref{SSK}, we recall its  Lax representation \cite{TL}
\[
L_t=[P, L],
\]
where
\begin{equation}\label{eqlax}
    L = \partial_{x}^3+\phi'\partial_{x}-\phi_x\mathcal{D}+\phi'_x,
\end{equation}
and
\begin{align*}
\quad P  =&\; 9\partial_{x}^5+15\phi'\partial_{x}^3-15\phi_x\mathcal{D}\partial_{x}^2+30\phi'_x\partial_{x}^2-15\phi_{xx}\mathcal{D}\partial_{x} \\
 &+(5\phi'^2+25\phi'_{xx})\partial_{x}-10(\phi_{xxx}+\phi_{x}\phi'){\cal D}+10\phi'_{xxx}+10\phi'\phi'_{x}.
\end{align*}
Thus, the corresponding linear spectral problem is
\begin{equation}\label{eigen}
L\varphi=\lambda\varphi.
\end{equation}

From above Lax pair, we see that it is natural to work with $\phi'$ rather than $\phi$. Therefore, we introduce $v=\phi'$ and rewrite \eqref{SSK} as
\begin{equation}\label{v_eq}
v_t+v_{xxxxx}+5vv_{xxx}+5v_xv_{xx}+5v_xv^2+5v'v'_{xx}=0.
\end{equation}

Our aim now is to find a Darboux transformation
for \eqref{eigen} and we will take the well adopted approach, namely gauge transformation approach. To this end, we first reformulate the linear spectral problem \eqref{eigen} into the matrix form and such reformulation is standard. Introducing $\Phi=(\varphi,\varphi_{x},\varphi_{xx},\varphi',\varphi'_{x},\varphi'_{xx})^T $,  we may rewrite (\ref{eigen}) in matrix form, that is,
\begin{eqnarray}\label{eq:4}
\Phi'=M\Phi,\quad M=\left(
    \begin{array}{cccccc}
      0 & 0 & 0 & 1 & 0 & 0 \\
      0 & 0 & 0 & 0 & 1 & 0 \\
      0 & 0 & 0 &0  & 0 & 1 \\
      0 & 1 & 0 & 0 & 0 & 0 \\
      0 & 0 & 1 &0  & 0 & 0 \\
      \lambda-v_x & -v & 0 & v' & 0 & 0 \\
    \end{array}
  \right).
\end{eqnarray}

Above matrix $M$  has both bosonic and fermionic variables as its entries, thus a super matrix.
As in \cite{xue},  we introduce an involution on the algebra of super matrices in the following way: given
any matrix $A=(a_{ij})_{i,j\in \mathbb{Z}}$, we define  $A^\dag=(a_{ij}^\dag)_{i,j\in \mathbb{Z}}$
and $a_{ij}^\dag=(-1)^{p(a_{ij})}a_{ij}$  with $p(a_{ij})$ denoting the parity of $a_{ij}$.

The idea of constructing a Darboux transformation for \eqref{eq:4} is to seek for a gauge matrix $T$ such that
\begin{equation}\label{eq:5}
  \Phi_{[1]}=T\Phi
\end{equation}
 solves
\begin{equation}\label{eq:6}
  \Phi'_{[1]}=M_{[1]}\Phi_{[1]},
\end{equation}
where $M_{[1]}$ is the matrix $M$ but with $v$ replaced by the new field variable $v_{[1]}$.
Now it is easy to see that \eqref{eq:5} and \eqref{eq:6} imply that the gauge matrix $T$ has to satisfy
\begin{equation}\
  T'+T^{\dag}M-M_{[1]}T=0,
\end{equation}
a crucial equation we have to find a proper solution for it. To this end, we take  the simplest ansatz, namely
 \[
 T=\lambda F+G, \; F=(f_{ij})_{6\times 6}, \;  G=(g_{ij})_{6\times 6}.
 \]

 A careful analysis and tedious calculations show that the matrices $F$ and $G$ may be taken as
\begin{equation}\label{feq}
F=\begin{pmatrix}1&0&0&0&0&0\\ a&1&0&0&0&0\\ 2a_{x}+\frac{1}{2}a^2&a&1&-a'&0&0\\
    0&0&0&1&0&0\\ 2a'&0&0&a&1&0\\
    3a'_x+2aa'&2a'&0&a_{x}+\frac{1}{2}a^2&a&1 \end{pmatrix},
\end{equation}
and
\begin{equation}\label{geq}
G=\left(\begin{array}{cccccc}
  g_{11} & \frac{1}{2}a^2 & a & a'_x-aa'&-a'&0\\[4pt]
  g_{21}&g_{22}& a_{x}+\frac{1}{2}a^2 &g_{24}&-aa'&-a'\\[4pt]
  g_{31}& g_{32} & g_{33} &  g_{34}& g_{35}&-a'_x-aa'\\[4pt]
  g_{41}&-a'_x+2aa' &2a' &  g_{44}&-a_{x}+\frac{1}{2}a^2&a\\[4pt]
  g_{51}& \frac{3}{2}a^2a' & a'_x+2aa' &  -3a'_xa'& g_{55}&\frac{1}{2}a^2\\[4pt]
  g_{61}& g_{62} & g_{63} &  g_{64}& g_{65}&g_{66}
\end{array}\right).
\end{equation}
where
\begin{align*}
g_{11}=&-\frac{1}{2}a_{x}a+\frac{1}{4}a^3+av-\frac{a'_xa'}{a}+\lambda_0,\; (\lambda_0\;  \text{is a constant}),\\
g_{21}=&-\frac{1}{2}a_{x}^2+\frac{1}{8}a^4+\lambda_0a+a_{x}v+\frac{1}{2}a^2v-a'_xa'+v'a',\\
g_{22}=&\;\frac{1}{2}a_{x}a+\frac{1}{4}a^3-\frac{a'_xa'}{a}+\lambda_0,\\
g_{24}=&\;\frac{a_xa'_x}{a}+\frac{1}{2}aa'_{x}-a_xa'-\frac{1}{2}a^2a'+2\lambda_0\frac{a'}{a},\\
g_{31}=&-\frac{3}{2}a_{x}^2a+\frac{3}{4}a_{x}a^3+3\lambda_0a_x-av^2+\frac{7}{2}aa_xv-\frac{1}{4}a^3v
       -2\lambda_0v-\frac{3}{2}aa'_xa' \\
       &-\frac{a'_xa'v}{a}+v'a'_x+av'a',\\
g_{32}=&\;\frac{3}{2}a_{x}a^2-\frac{1}{2}a^2v-3a'_xa',\\
g_{33}=&\;3a_{x}a-av-2\frac{a'_xa'}{a}-\lambda_0,\\
g_{34}=&\;3a_xa'_x-a'_xv-3aa_{x}a'+aa'v+3\lambda_0a',\\
g_{35}=&\;\frac{a_xa'_x}{a}-\frac{1}{2}aa'_{x}-2a_xa'-\frac{1}{2}a^2a'+2\lambda_0\frac{a'}{a}+a'v,\\
g_{41}=&\;\frac{a'_xa_x}{a}-\frac{1}{2}aa'_{x}-2a_xa'+a^2a'+2\lambda_0\frac{a'}{a}+2a'v+v'a,\\
g_{44}=&-2\frac{a'_xa'}{a}-\lambda_0,
\\
g_{51}=&\;a'_xv-\frac{3}{2}aa_xa'+\frac{3}{4}a^3a'+2aa'v+3\lambda_0a'+\frac{1}{2}v'a^2,\\
g_{55}=&-\frac{1}{2}a_{x}a+\frac{1}{4}a^3-\frac{a'_xa'}{a}+\lambda_0,\\
%\end{align*}
%\begin{align*}
g_{61}=&-\frac{3}{2}aa_xa'_x+\frac{3}{4}a^3a'_x+\frac{a_xa'_xv}{a}+\frac{7}{2}aa'_xv+3\lambda_0a'_x-\frac{3}{2}a_x^2a'+\frac{3}{8}a^4a'  \\
       &+3\lambda_0aa'+2a_xa'v+a^2a'v+2\lambda_0\frac{a'v}{a}+aa_xv'-av'v,\\
g_{62}=&\;\frac{3}{2}a^2a'_x+\frac{3}{2}aa_xa'+\frac{3}{4}a^3a'+3\lambda_0a'-\frac{1}{2}v'a^2,\\
g_{63}=&\;\frac{a_xa'_x}{a}+\frac{7}{2}aa'_{x}+2a_xa'+a^2a'+2\lambda_0\frac{a'}{a}-v'a,\\
g_{64}=&-3\frac{a_xa'_xa'}{a}-\frac{9}{2}aa'_xa'-v'a'_x+av'a',\\
g_{65}=&-\frac{1}{2}a_x^2+\frac{1}{8}a^4+\lambda_0a-4a'_xa'+v'a',\\
g_{66}=&\;g_{22}.
\end{align*}
It is noticed that all the entries of the Darboux matrix $G$ are represented in term of the field variable $v$ or $\phi$, an auxiliary (bosonic) variable $a$ and their derivatives. In addition, the single auxiliary variable $a$ satisfies the following equation
\begin{equation}\label{bt}
   a_{xx}=-\frac{1}{4}a^3+\frac{3}{2}aa_x-2\lambda_0-av+\frac{a'a'_x}{a},
\end{equation}
and the transformation between field variables reads as
 \begin{equation}\label{dtpotential}
   v_{[1]}=v-3a_x.
\end{equation}

The equation \eqref{dtpotential} may be used to eliminate the auxiliary variable $a$ in \eqref{bt} and in this way  a B\"acklund transformation (spatial part) for the SSK equation \eqref{SSK} may be obtained.

\bigskip
\noindent
{\bf Remark}: For above B\"acklund transformation, we may take its bosonic limit and find
\begin{equation}\label{skbt}
   a_{xx}=-\frac{1}{4}a^3+\frac{3}{2}aa_x-2\lambda_0-au,
\end{equation}
and
 \begin{equation}\label{skdtpotential}
   u_{[1]}=u-3a_x.
\end{equation}
It is easy to see that this is nothing but the B\"acklund transformation of SK equation \eqref{SK}, first appeared in \cite{satsuma-k,dodd-g}.

\bigskip

While we have worked out a B\"acklund transformation for the SSK equation, the Darboux matrix $T$ is implicit in the sense that it depends on  $a$  and it is desirable to relate it the solutions of the linear spectral problem \eqref{eq:4} in such way that the Darboux matrix $T$ may take an explicit form.  Thus, we consider the kernel of the Darboux matrix $T$ and  take the particular solution $\Phi_0=(\varphi_0,\varphi_{0x},\varphi_{0xx},\varphi'_0,\varphi'_{0x},\varphi'_{0xx})^{\texttt{T}} $ of \eqref{eq:4} at $\lambda=\lambda_0$ such that $T\Phi_0=0$. Then  we  find
\begin{equation}\label{dt}
   a=-\frac{2\lambda_{0}\varphi_{0}}{v\varphi_{0}+\varphi_{0xx}}+\frac{2\lambda_{0}\varphi'_{0x}\varphi'_{0}}{[v\varphi_{0}+\varphi_{0xx}]^2},
\end{equation}
where $\varphi_{0}$ is a bosonic function.

Now summarizing above discussions, we have
\begin{prop}
Let $\varphi_{0}$ is a bosonic solution of the linear spectral problem \eqref{eigen} at $\lambda=\lambda_0$. Let the matrices $F$ and $G$  be given by \eqref{feq} and \eqref{geq} with the quantity $a$ given by \eqref{dt}. Then $T=\lambda F+G$ is a Darboux matrix for the linear spectral problem \eqref{eq:4}. The transformation for the   field variables is given by \eqref{skdtpotential}.
\end{prop}

It is interesting to note that the scalar version of the Darboux transformation may be obtained as follows
\begin{equation}\label{eq1}
\begin{aligned}
\varphi_{[1]}=&a\varphi_{xx}-a'\varphi'_x+\frac{1}{2}a^2\varphi_x+(a'_x-aa')\varphi'\\\nonumber
&+
\left[\lambda+\frac{1}{4}a^3-\frac{1}{2}aa_x+\lambda_0+av
+\frac{a'a'_x}{a}\right]\varphi.
\end{aligned}\end{equation}

As a first application, we now employ the Darboux transformation  to build the 1-soliton solution for the SSK system (\ref{SSK}). We begin with the vacuum seed $v=0$ and consider the corresponding linear problem
\begin{align}\label{0lax}
&\varphi_{0xxx}=\lambda_{0}\varphi_{0},\;\;\;\; \varphi_{0t}=9\lambda_{0}\varphi_{0xx}=9\varphi_{0,xxxxx}.
\end{align}
Assuming $\lambda_{0}=\frac{i}{3\sqrt{3}}k^3$ $(i=\sqrt{-1},\; k\in \mathbb{R})$, we easily find that
\[
 \varphi_{0}=e^{px+9p^5t+\sqrt{3}/3i\theta\zeta}(1+e^{kx-k^5t-2\pi i/3+c}),
\]
where  $p=(-\frac{1}{2}+\frac{\sqrt{3}}{6}i)k$, $\zeta$ and $c$ are  arbitrary fermionic constant and  bosonic constant respectively,
solves the system \eqref{0lax}.
Now from \eqref{dt}, we obtain
\[
a=-k\left(\frac{\sqrt{3}}{3}i+\tanh\eta\right),\; \eta=\frac{1}{2}(kx-k^5t+c+\theta\zeta),
\]
and substituting them into \eqref{dtpotential} leads to the following 1-soliton solution of the SSK equation
\[
v=\frac{3}{2}k^2\mbox{sech}^2\eta.
\]
It is noted  that the bosonic part of the solution is just the 1-soliton solution of the SK equation \cite{sk, caudrey-dg}.
%Furthermore by  specifying the %constants as follows
%\begin{align*}
%c_{1}=1-\sqrt{3}i,\;\;\;\; c_{2}=1+%\sqrt{3}i,\;\;\;\; c_{3}=0,\;\;\;\; \zeta_{1}=\zeta_{2}=\zeta,
%\end{align*}
% we obtain
%\begin{equation}\label{1-soli}
%  v_{[1]}=\frac{72k_{0}^{2}%e^{\eta_{[1]}-\eta_{[2]}}}%{(1+e^{\eta_{[1]}-\eta_{[2]}})^2}
%           +\frac{72k_{0}^{2}\theta%\zeta e^{\eta_{[1]}-\eta_{[2]}}(1-%e^{\eta_{[1]}-\eta_{[2]}})}%{(1+e^{\eta_{[1]}-\eta_{[2]}})^3},
%\end{equation}
%where $\eta_{[1]}=px+9p^5t$, $\eta_{[2]}=rx+9r^5t$.

%Introducing  $k={2\sqrt{3}}k_0$, $\eta=\eta_{{[1]}}-\eta_{{[2]}}$, we rewrite \eqref{1-soli} as
%\begin{equation*}\label{}
%   v_{[1]}=\frac{6k^{2}e^{\eta}}%{(1+e^{\eta})^2}
%           +\frac{6k^{2}\theta\zeta %e^{\eta}(1-e^{\eta})}{(1+e^{\eta})^3},
%\end{equation*}
%which is  the one-soliton solution of %SSK equation.

\section{Nonlinear Superposition Formula}
In the last section we constructed the B\"acklund transformation and Darboux transformation for the SSK equation. It was shown that these transformations may be used to build solutions for the SSK equation. However,   B\"acklund transformation  itself is a system of differential equations, therefore it may not be easy  to solve for more general seed solutions. The usual  way to get over this difficulty is to derive the corresponding nonlinear superposition formula, which we now look for.

It turns out that a convenient way is  to work with the potential form of the equation, so we introduce $\phi=3w'$ or $v=3w_x$ and convert the SSK equation \eqref{SSK}  into
\begin{equation}\label{pSSK}
   w_{t}+w_{xxxxx}+15w_{xxx}w_{x}+15w_{x}^3+15w'_{x}w'_{xx}=0.
\end{equation}
Now we suppose that
$w$ is an arbitrary solution of the potential SSK \eqref{pSSK} and $\lambda_{j}\;(j=1,2)$ are arbitrary constants taken as B\"acklund parameters, then we may perform Darboux transformation $\Phi_{[j]}=T|_{\lambda=\lambda_{j}}\Phi$ and find new solution  $w_{j}$. Namely, we consider a pair of Darboux transformations
\begin{align}
&\Phi_{[1]}=T_{[1]}\Phi, \;\;\; T_{[1]}\equiv T|_{\lambda_0=\lambda_{1}, a=a_1}\label{phi1},&\\
&\Phi_{[2]}=T_{[2]}\Phi, \;\;\; T_{[2]}\equiv T|_{\lambda_0=\lambda_{2}, a=a_2},&\label{phi2}
\end{align}
where
\[
  a_{1}=w-w_{1},\quad a_{2}=w-w_{2}.
\]
Then with the help of  the Bianchi's permutability theorem,  represented schematically by the diagram below
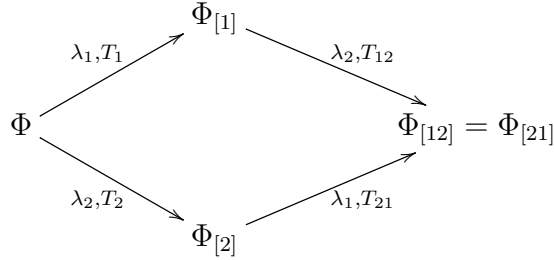
\begin{displaymath}
     \xymatrix{
                                & &\Phi_{[1]}\ar[drr]^{\lambda_2, T_{12}}&&\\
         \Phi\ar[urr]^{\lambda_1, T_1} \ar[drr]_{\lambda_2, T_2}  &  &&&\Phi_{[12]}= \Phi_{[21]}
         \\
                             && \Phi_{[2]}\ar[urr]_{\lambda_1, T_{21}}&& }
\end{displaymath}
we obtain
\begin{equation}\label{eq5}
  T_{[12]} T_{[1]} =T_{[21]} T_{[2]},
\end{equation}
where
\[
 T_{[12]}\equiv T|_{\lambda_0=\lambda_{2}, a=a_{12}}, \quad T_{[21]}\equiv T|_{\lambda_0=\lambda_{1}, a=a_{21}},  \quad
  a_{12}=w_{1}-w_{12},\quad a_{21}=w_{2}-w_{21}.
\]

After some cumbersome calculations, we find, from \eqref{eq5},  the following nonlinear superposition formula
\begin{align}\label{nsf}
   w_{12}=&\;w+\frac{4(\lambda_{1}+\lambda_{2})(a_1-a_2)}{\Delta_1} +\frac{16(\lambda_{1}+\lambda_{2})}{\Delta_1^2}\bigg[(a_2-a_1)a'_{1}a'_{2}\\\nonumber
 &\left.  +a_{1}^{2}
 \left(\frac{a'_{1x}}{a_1}-\frac{a'_{2x}}{a_2}\right)\left(\frac{a_2}{a_1}\right)'\right],
\end{align}
where
\[\Delta_1=4(\lambda_{2}-\lambda_{1})+a_1a_2(a_2-a_1)+2a_2a_{1x}-2a_1a_{2x}.
\]
We now employ above nonlinear superposition formula and build a 2-soliton solution to the SSK equation. To this end, for arbitrary bosonic constants $k_j$ and fermionic constants $\zeta_j$ ($j=1,2$) we take
\[w=0,\;\; \lambda_j=\frac{i}{3\sqrt{3}}k_j^3,
\]
and
\[
 a_{1}=-k_1\left(\frac{i}{\sqrt{3}}+\tanh({\eta}_{1}-\frac{1}{2}ix_0)\right),\;\;  a_{2}=-k_2\left(\frac{i}{\sqrt{3}}+\tanh({\eta}_{2}+\frac{1}{2}ix_0)\right),
 \]
where
\[
{\eta}_{j}=\frac{1}{2}(k_j x-k_j^5t-c_j+\theta\zeta_j),
%\;\; \tilde{\eta}_{2}=\frac{1}{2}%(k_2x-k_2^5t+x_1i-\pi i+\theta\zeta_2),\;
\;x_0=\arctan\left(\frac{\sqrt{3}k_1k_2}{k_2^2-k_1^2}\right).
\]
%and $k_\ell\in \mathbb{R}$,  $k_j$ and $ \zeta_j \; (j=1,2)$ are arbitrary real constants and  fermionic constants respectively.
Then from \eqref{nsf}, by some tedious  calculations we obtain
%\begin{align}\label{p2soliton}
%  w_{12}=&\;\frac{4c_0^3(k_1^3+k_2^3)
%(c_0(k_1-k_2)+k_2\coth\tilde{\eta}_{2}-k_1\tanh\tilde{\eta}_{1})}{\Delta_1}\nonumber\\
% &+\frac{4c_0^4k_1k_2(k_1-k_2)
%(k_1^3+k_1^3)\mbox{sech}^{2}\tilde{\eta}%_{1}\mbox{csch}^{2}\tilde{\eta}_{2}}
%{\Delta_{1}^2}(\zeta_1+k_1\theta)
%(\zeta_2+k_2\theta),
%\end{align}
%where
%\begin{align*}
%\Delta_1=&\;4c_0^3(k_2^3-k_1^3)+k_1k_{2}^2(c_0-\tanh\tilde{\eta}_{1})(1+c_{0}^2-2c_0\coth\tilde{\eta}_{2})  \\
%        &-k_1^2k_{2}(c_0-\coth\tilde{\eta}_{2})(1+c_{0}^2-2c_0\tanh\tilde{\eta}_{1}).
%\end{align*}
%To see that \eqref{p2soliton} gives a two-soliton solution, we  rewrite it as
\begin{align}\label{2soliton}
  v_{12}= &\;\frac{3\big[b_1+k_2^2b_0\cosh(2{\eta}_{1})+k_1^2b_0\cosh(2\eta_2)\big]}{\big[\cosh(\eta_1+\eta_2)+b_0\cosh(\eta_1-\eta_2)\big]^2}\nonumber\\
  &-\frac{b_2\big[b_3\sinh(\eta_1+\eta_2)+b_0\sinh(\eta_1-\eta_2)\big]}{\big[\cosh(\eta_1+\eta_2)+b_0\cosh(\eta_1-\eta_2)\big]^3}(\zeta_1+\theta k_1)(\zeta_2+\theta k_2),
\end{align}
where
\begin{align*}
%{\eta}_{1}=&\frac{1}{2}(k_1x-k_1^5t+
%\theta\zeta_1),\;\; {\eta}_{2}=\frac{1}
%{2}(k_2x-k_2^5t+\theta\zeta_2),\\
b_0=&\sqrt{\frac{(k_1+k_2)^3}{k_1^3+k_2^3}\frac{(k_1^2+k_1k_2+k_2^2)}{(k_1-k_2)^2}},&
b_{1}=&\frac{(k_1+k_2)^3}{k_1^3+k_2^3}(k_1^2+k_2^2),\\
b_{2}=&\frac{9k_1k_2(k_1+k_2)^2}{k_1^3+k_2^3},&
b_{3}=&\frac{k_1+k_2}{k_1-k_2}.
\end{align*}
Thus, $v_{12}$ given by \eqref{2soliton} is a 2-soliton solution of the SSK equation. It is easy to check that the  bosonic  limit of the above solution is nothing but the 2-soliton solution  of the SK equation.

%\noindent
%{\bf Remark}:
We remark that our nonlinear superposition formula is of differential-algebraic type which may serve as an effective way to build more solutions. Also, by taking the bosonic limit we may have the following nonlinear superposition formula for the SK equation
\[
 w_{12}=w+\frac{4(\lambda_{1}+\lambda_{2})(a_1-a_2)}{\Delta_1},
\]
which should be compared with \cite{hu-l} (see also \cite{musette}).
%, last formula does not involve with %integrals, this is the advantage of using %potential variables.

\section{Discretizations and continuous limits}
It is well known that in addition to allowing one to construct solutions of nonlinear systems, B\"acklund transformations and the associated nonlinear superposition formulae often supply new integrable systems of both continuous and discrete types. Next, we will show that this is the case for our B\"acklund transformation and nonlinear superposition formula derived above.
\subsection{Discrete systems}

We first consider B\"acklund transformation. To do so, we write out \eqref{bt}, using the potential variables  $w_{x}$, $w_{1x}$, and $\lambda_0=\lambda_1$, as follows
%the B\"{a}cklund transformation \eqref{bt} is %split into the system
\begin{equation}\label{pbt}
  (w-w_{1})_{xx}=-\frac{1}{4}(w-w_{1})^3+\frac{3}{2}(w_{1}-w)(w_{1}+w)_x
      -2\lambda_{1}+\frac{(w'-w'_{1})(w'-w'_{1})_x}{w-w_{1}}.
\end{equation}
Let $\lambda_{1}=-\frac{1}{8}p_{1}^3$, $w\rightarrow w+p_1$,
we have
\begin{align}
 (w_{1}-w)_{xx}=&-\frac{3}{2}(w_{1}-w-p_1)(w_{1}+w)_x-\frac{1}{4}(w_{1}-w)^3  \nonumber \\
 &+\frac{3}{4}(w_{1}-w)^{2}p_{1}  -\frac{3}{4}(w_{1}-w)p_{1}^{2}+\frac{(w'_{1}-w')(w'_{1}-w')_x}{w_{1}-w-p_1}.
\end{align}
Define
\begin{eqnarray}\nonumber
   w\equiv w_{n}(x),\quad \,  w_{1}\equiv w_{n+1}(x).
\end{eqnarray}
We finally get
\begin{align}\label{btdis}
(w_{n+1}-w_{n})_{xx}=&-\frac{3}{2}(w_{n+1}-w_{n}-p_1)(w_{n+1}-w_{n})_x-\frac{1}{4}(w_{n+1}-w_{n})^3  \nonumber\\
    & +\frac{3}{4}(w_{n+1}-w_{n})^{2}p_{1}-\frac{3}{4}(w_{n+1}-w_{n})p_{1}^{2}\nonumber\\
    &+\frac{(w'_{n+1}-w'_{n})(w'_{n+1}-w'_{n})_x}{w_{n+1}-w_{n}-p_1}.
\end{align}
It is a differential-difference system.

For the nonlinear superposition formula, let $\lambda_{1}=-\frac{1}{8}p_{1}^3$,  define for any field variable $w$
\[
w\equiv w_{n,m},\quad \;w_{1}\equiv w_{n+1,m},\;\quad w_{2}\equiv w_{n,m+1},\;\quad w_{12}\equiv w_{n+1,m+1},
\]
and replace $w_{n,m}$  by $w_{n,m}-np_1-mp_2$, from \eqref{nsf} we obtain
\begin{equation}\label{nsfdis}
\frac{w_{n+1,m+1}-w_{n,m}-p_1-p_2}{p_1^3+p_2^3}=\frac{S_1-S_2}{\Delta_2}-\frac{8}{\Delta_2^2}\left[(S_1-S_2)S'_1S'_2+(S_2S'_{1,x}-S_1S'_{2,x})\left(\ln\frac{S_2}{S_1}\right)'\right],
\end{equation}
where
$S_1=w_{n+1,m}-w_{n,m}-p_1, S_2=w_{n,m+1}-w_{n,m}-p_2$ and
\[
\Delta_2=p_1^3-p_2^3+2S_1S_2\left[(S_1-S_2)+2\left(\ln\frac{S_1}{S_2}\right)_x
\right].
\]
It is noted that \eqref{nsfdis} is a differential-partial difference system.

Taking the bosonic limits of \eqref{btdis} and \eqref{nsfdis}, we find two differential-difference systems as follows
%{\color{red} If considered  the bosonic limit of the discrete systems \eqref{btdis} and \eqref{nsfdis}, we can get the following two ordinary discrete SK systems  respectively

\begin{align*}\label{classicalbtdis}
(w_{n+1}-w_{n})_{xx}=&-\frac{3}{2}(w_{n+1}-w_{n}-p_1)(w_{n+1}-w_{n})_x-\frac{1}{4}(w_{n+1}-w_{n})^3  \nonumber\\
    & +\frac{3}{4}(w_{n+1}-w_{n})^{2}p_{1}-\frac{3}{4}(w_{n+1}-w_{n})p_{1}^{2},
\end{align*}
and
\begin{equation*}\label{classicalnsfdis}
 w_{n+1,m+1}=w_{n,m}+p_1+p_2+\frac{(p_1^3+p_2^3)(S_1-S_2)}{\Delta_2},
\end{equation*}
they are different from the known semi-discrete versions of the SK equation (cf. \cite{tsujimoto}\cite{adler}\cite{hu-2})

%\begin{eqnarray}
%\Delta_2=p_1^3-p_2^3+2(v_{n+1,m}-v_{n,m}-p_1)%%%(v_{n,m+1}-v_{n,m}-p_2)(v_{n+1,m}-v_{n,m+1}-%p_1+p_2)  \nonumber\\
%+4(v_{n,m+1}-v_{n,m}-p_2)(v_{n+1,m}-%v_{n,m})_x-4(v_{n+1,m}-v_{n,m}-p_1)(v_{n,m+1}-%v_{n,m})_x .
%\end{eqnarray}

\subsection{Continuum limits}
As a final part, we relate the semi-discrete systems obtained last subsection to the SSK equation. We will show that by taking proper continuum limits both (\ref{btdis}) and  (\ref{nsfdis}) go to the potential SSK equation \eqref{pSSK}.
%In the last section, with the help of the B%\"acklund transformation and the related %nonlinear superposition formulae, we obtained %two discrete systems . It is interesting to %identify these discrete systems and analyze  %their continuum limits \cite{frank}. We will %show that the obtained systems  are nothing %but the discrete versions of the potential SSK %(\ref{pSSK}).

For the differential-difference system (\ref{btdis}), we introduce the new continuous variable $\tau$ as
\begin{align*}
w_{n}(x)\equiv w(x,\tau),
\end{align*}
then
\begin{align*}
w_{n+1}(x)\equiv w\left(x,\tau+\frac{1}{p_1}\right)
\end{align*}
may be expanded in $\frac{1}{p_1}$, and defining  a new independent variable  $t$ in term of $\tau$ and $x$ such that
\begin{align*}
\partial_\tau=4\partial_x+\frac{64}{45{p_1}^4}\partial_t,
\end{align*}
in the continuous limit up to terms of order $\frac{1}{p_{1}^3}$, we find
\begin{equation}\label{eq10}
   w_{t}+w_{xxxxx}+15w_{xxx}w_{x}+15w_{x}^3+15w'_{x}w'_{xx}=0,
\end{equation}
which is  the potential  form of  the SSK equation (\ref{pSSK}).

For the differential-partial difference system (\ref{nsfdis}), we consider the so-called straight continuum limit \cite{frank}.
Thus, we assume
\begin{align*}
w_{n,m}\equiv w_{n}(x)\equiv w_{n}(\frac{4m}{p_2}).
\end{align*}
For $\frac{1}{p_2}$ small, we take the following Taylor series expansions
\begin{align*}
w_{n,m+1}=w_{n}\left(x+\frac{4}{p_2}\right)&=w_{n}+\frac{4}{p_2}w_{n,x}+\frac{8}{p_2^2}w_{n,xx}+O\left(\frac{1}{{p_2}^3}\right),\\
w_{n+1,m+1}=w_{n+1}\left(x+\frac{4}{p_2}\right)&=w_{n+1}+\frac{4}{p_2}w_{n+1,x}+\frac{8}{p_2^2}w_{n+1,xx}+O\left(\frac{1}{{p_2}^3}\right),
\end{align*}
plugging above equations into (\ref{nsfdis}), then the leading terms yield  the system (\ref{btdis}). Therefore, we may say that the  (semi-)discrete system (\ref{nsfdis}) is a discrete version of the potential SSK system.

%The bosonic part of equation (\ref{eq10}) is %the potential versions of  (\ref{SK}) read as
%\begin{equation}\label{pSK}
%   w_{t}+w_{xxxxx}+15w_{xxx}w_{x}+15w_{x}^3=0.
%\end{equation}
%Naturally,  the bosonic parts of systems %(\ref{btdis}) and  (\ref{nsfdis}) are nothing %but the discrete versions of the potential SK %(\ref{pSK}).
Of course, we may follow \cite{frank} and study other continuum limits such as skew continuum limit or full continuum limit for the system (\ref{nsfdis}), but such calculations will not be given here since they are somewhat cumbersome.
% we will not present the results  here since %these can be done thorough  tedious but %straightforward calculations.

\bigskip

\noindent
{\bf Acknowledgement}\\
We should like to thank the anonymous referees for their suggestive comments. The work is supported by the National Natural Science Foundation of China (grant numbers: 11271366, 11331008 and 11501312), Zhejiang Provincial Natural Science Foundation of China (grant number: LQ15A010002) and the Fundamental Research Funds for Central Universities.

\end{document}